\documentstyle[a4,11pt,epsf,graphicx,twoside]{article}

\begin{document}
\setcounter{page}{1}
\title{READOUT ELECTRONICS FOR THE CALICE ECAL AND TILE HCAL}
\author{P. D. DAUNCEY\thanks{e-mail address: \tt P.Dauncey@imperial.ac.uk}
        \ representing the CALICE-UK Collaboration
\\
\\
       {\it Physics Department, Imperial College London, UK}
}
\date{}
\maketitle
\begin{abstract}
The aims of the CALICE test beam program are presented.
The proposed electronics readout system for the CALICE ECAL
is described. It is a purpose-built VME-based, unbuffered system.
\end{abstract}

\section{The CALICE Collaboration}
The CALICE collaboration~\cite{calice} is a group of 140 people from
24 institutes in Europe, US and Asia. The collaboration is planning
a beam test program starting in 2004 to study both electromagnetic
(ECAL) and hadronic (HCAL) calorimeters for a future linear collider (LC).
The ECAL will be a
silicon-tungsten sandwich sampling calorimeter~\cite{ecal} and
there are various options under investigation for the HCAL, namely
using tile scintillators with analogue readout~\cite{thcal}
or RPCs or GEMs with digital
readout~\cite{dhcal} as the sensors. The main driver for these
choices is the requirement of fine granularity for the ``energy flow''
algortihms used to obtain the jet resolutions necessary for LC 
physics~\cite{eflow}.

The beam test program will test the ECAL and all versions of the HCAL
in an integrated data acquisition system so the data from both
calorimeters can be analysed together.
It is planned to do a systematic study of the dependence of the
shower structure on particle energy, type and incident angle, as well
as adding a ``preshower'' in front of the calorimeter to generate
narrow cones of particles to simulate jets.
The total number of different configurations which will be tested
is of order $10^2$, and to see the fine detail required, of order
$10^6$ events is estimated to be needed for each configuration.

\section{ECAL electronics}

The CALICE beam test ECAL will consist of 30 layers of silicon and
tungsten. Each silicon layer comprises nine wafers, each being a
$6 \times 6$ array of diode pads and each of these is a readout channel.
This gives 324 channels per layer, or 9720 channels in total. These
need to be read for every event, preferably without any threshold or
other data suppression so as to allow pedestal and noise studies
offline.

The signals expected in the diodes range from the energy deposited by
one minimum ionising
particle (MIP) to around 1000 MIPs in the core of high energy
electron showers, a range of 10 bits. The intrinsic noise from the
on-detector preamplifier~\cite{vfe} is expected to be equivalent to
around 0.1 MIP or less. It is of interest to measure this noise level,
so the dynamic range required is 14 bits in total. At the highest
energies of around 1000 MIPs, 
the resolution is degraded if the precision is below a
MIP, so 10 bits of precision are required.
The on-detector preamplifier has a CR-RC shaper with a peaking
time of 180ns. This defines the trigger latency needed to ensure
the waveform is sampled at this peak. A trigger jitter of 10ns is
acceptable; this leads to a maximum shift below the peak of 0.1\%,
which is within the 10 bits precision requirement.

\begin{figure}
\includegraphics[height=5.0cm]{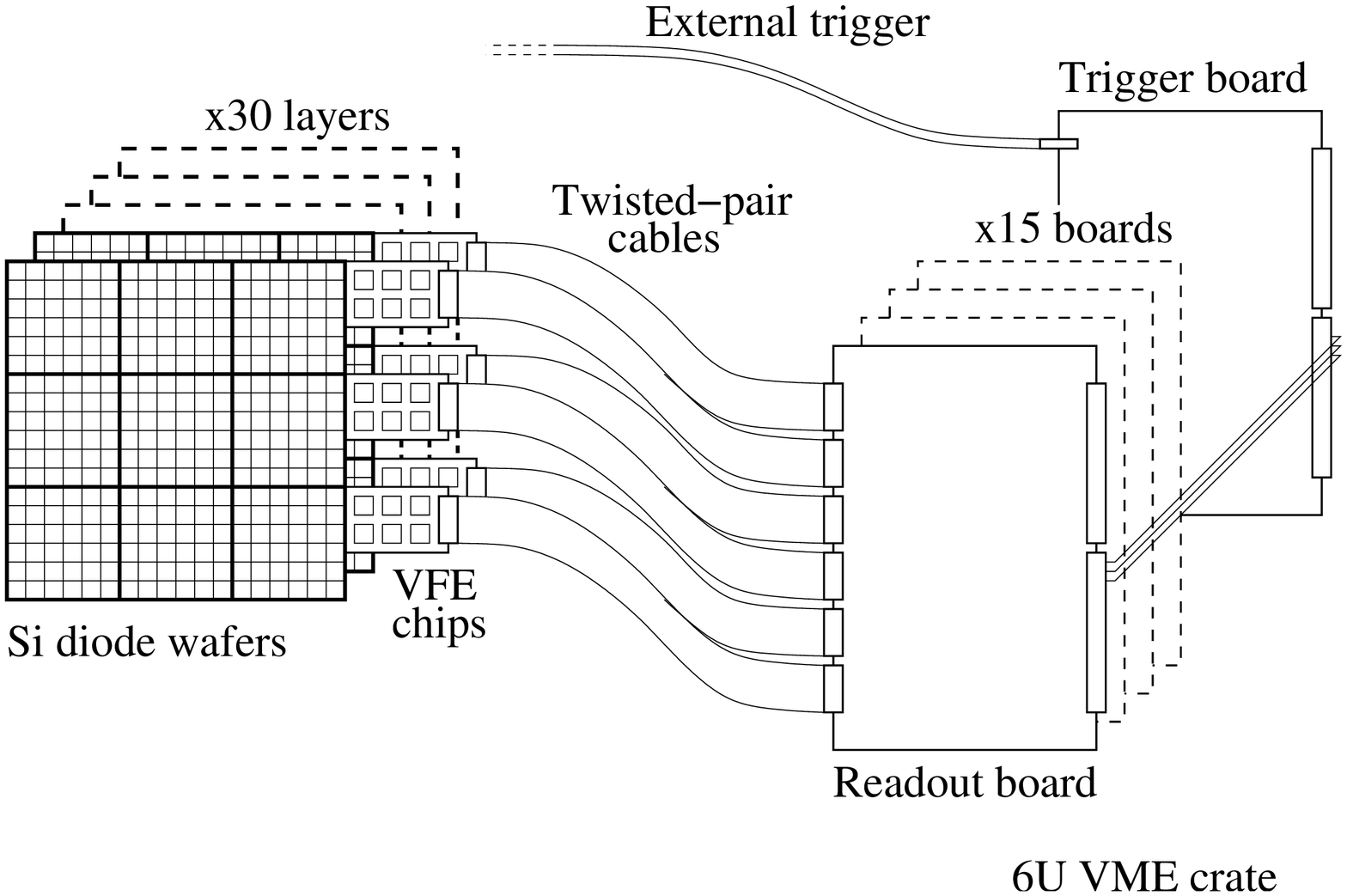}
\hfill
\includegraphics[height=5.7cm]{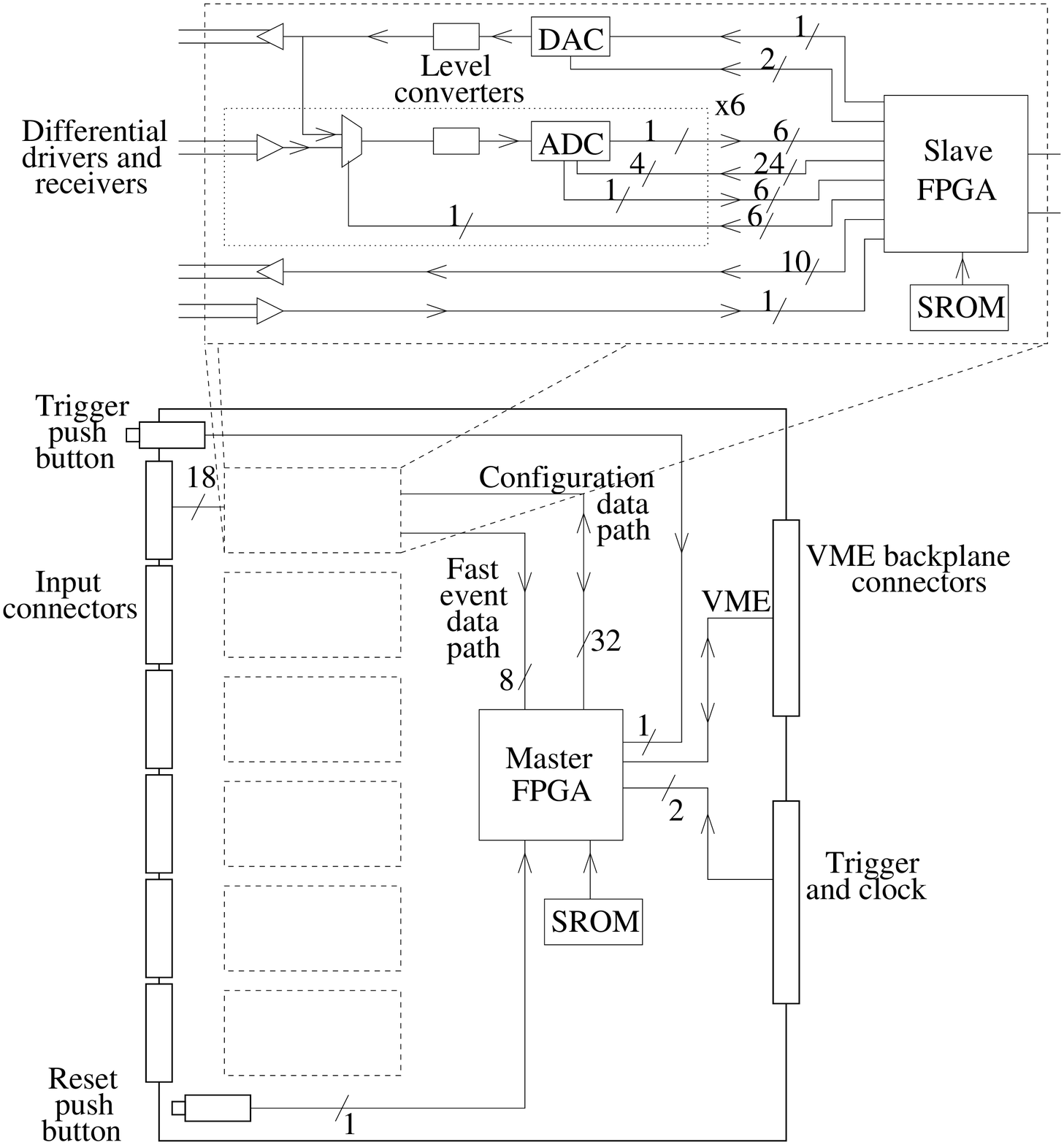}
\caption{Overview of the electronics system. The left figure shows the
layout of electronics boards in the VME crate. The right figure shows
the functional layout of the readout board.
\label{fig:overview}}
\end{figure}

The proposed system~\cite{caliceuk} is shown in figure~\ref{fig:overview}.
It consists of 15
identical readout boards and a single trigger board. The readout
boards handle the on-detector preamplifier signals and all digitisation
for two layers of the ECAL, or 648 channels. These signals
are received on six cables per readout board. Each cable connects to
an electrically independent on-detector board which holds the
preamplifier and which handles three silicon wafers of 108 channels.

The trigger board
is a simple board to prevent further triggers until readout is complete
and to distribute the trigger and the system clock across the backplane
to the readout boards. It contains very few components and will not be
described further below.

An schematic diagram of the readout board is also shown 
in figure~\ref{fig:overview}.
The board is controlled by a ``master FPGA'' which handles the VME
interface and distributes the clock and trigger from the backplane across the 
board. It interacts with the ``slave FPGAs'' by distributing
configuration data (written to the board via VME) and collects the event
data prior to VME readout.

There is one slave FPGA for each cable from the on-detector electronics
and so six per readout board. Each operates independently of the others
although they contain identical firmware so,
in standard operation, they would effectively be
synchronised together. The slave FPGAs send the control 
and timing signals needed by the on-detector electronics and receive
six analogue signal lines. The 108 channels are multiplexed onto these
six lines, with 18 channels per line. The slave FPGAs control six
16-bit, 500kHz ADCs, one per signal line. The extra bits give
robustness against loss due to non-optimal range matching. The
ADC speed results in a time to digitise all 18 channels of less
than 50$\mu$s, or only 5\% of the allowed 1ms event time.
The trigger is also distributed to the on-detector electronics via
the slave FPGAs, rounded to a 100 MHz clock, giving the required
10ns maximum jitter.

To save any parallel development of similar electronics for the
tile HCAL option, the on-detector electronics for this system is
attempting to become similar enough to the ECAL such that the
same readout boards could be used for both.
There will be around 1500 channels in the tile scintillator HCAL.
The 16-bit ADC would be more than ample for the range and precision
required and would result in 3kBytes per event.
The number of channels multiplexed per board for the HCAL is also
not known, so it is not clear how many extra readout boards would
need to be fabricated.
However, it is clear that a common solution would save significant
effort and is being actively pursued by the groups involved.

The maximum rate of data readout over VME is around 30MBytes/s.
For a desired 1kHz rate, this implies the event size needs to be below
30kBytes. There will be no threshold suppression of the data on
the readout board, so the ECAL data size will be 19kBytes per
event. Both options of the HCAL will be around 3kBytes and the size
of any beam monitoring, particle idenitification and trigger data are
uncertain, but likely to be around 1kByte. Hence, the total event
size should be around 25kBytes. This makes a 1kHz peak rate tight but
not impossible. The readout board VME interface will be optimised
for speed, using DMA transfers and asynchronous VME access. A fallback
solution will be to split the system between two or more crates, with two
VME-PCI bus convertors.

\end{document}